\documentclass[aps,reprint]{revtex4-2}

\usepackage{graphicx}
\usepackage{bm}
\usepackage{amsmath}
\usepackage{amssymb}
\usepackage{physics}

\usepackage{color}

\usepackage{hyperref}
\hypersetup{
    colorlinks=true,
    linkcolor=blue,
    filecolor=magenta,      
    urlcolor=blue,
    }

\begin{document}

\title{Excitation transfer in disordered spin chains with long-range exchange interactions}

\author{Nikolaos E. Palaiodimopoulos}
\affiliation{Institute of Electronic Structure and Laser, FORTH, GR-70013 Heraklion, Greece}

\author{David Petrosyan}
\affiliation{Institute of Electronic Structure and Laser, FORTH, GR-70013 Heraklion, Greece}

\author{Maximilian Kiefer-Emmanouilidis}
\affiliation{Department of Physics and Research Center OPTIMAS,
University of Kaiserslautern, D-67663 Kaiserslautern, Germany}

\author{Gershon Kurizki}
\affiliation{Department of Chemical and Biological Physics, Weizmann Institute of Science, Rehovot 7610001, Israel}

\begin{abstract}
We examine spin excitation or polarization transfer via long-range interacting spin chains with diagonal and off-diagonal disorder. 
To this end, we determine the mean localization length of the single-excitation eigenstates of the chain for various strengths of the disorder. 
We then identify the energy eigenstates of the system with large localization length and sufficient support at the chain boundaries that are suitable to transfer an excitation between the sender and receiver spins connected to the opposite ends of the chain. 
We quantify the performance of two transfer schemes involving weak static couplings of the sender and receiver spins to the chain, and time-dependent couplings realizing stimulated adiabatic passage of the excitation via the intermediate eigenstates of the chain which exhibits improved performance.
\end{abstract}

\maketitle

\newpage

\section{Introduction}

Excitation or polarization transfer in interacting few- and many-body quantum systems plays a key role in many brunches of science and technology, 
ranging from photosynthesis, where photon energy is transferred from a light-absorbing center to a reaction center via collections of near-resonant two-level systems (spins) \cite{mirkovic2017light}, 
nuclear magnetic resonance of large molecules involving many interacting spins \cite{foster2007solution}, 
or quantum state transfer in various spin chains realized, e.g., 
by dopants in solids \cite{randall2021many, lake2021direct,alvarez2015localization}, 
arrays of polar molecules \cite{gulli2019macroscopic,yan2013observation}, 
superconducting qubits \cite{dalmonte2015realizing}, 
ions in traps \cite{jurcevic2014quasiparticle,britton2012engineered} or 
Rydberg atoms in microtraps \cite{browaeys2016experimental}.
Whereas spin chains are commonly described in the nearest-neighbour approximation, experimentally relevant systems often possess long-range exchange interactions scaling with distance $r$ as $J \sim 1/r^{\nu}$ with the resonant dipole-dipole interaction, $\nu=3$, being most frequently the case.

Many of such systems are inherently disordered. 
Diagonal disorder leads to exponential (Anderson) localization of all the eigenstates of one-dimensional systems \cite{mott1961theory,anderson1958absence,gogolin1982}, which would suppresses excitation transfer in sufficiently long spin chains. 
Off-diagonal disorder also leads to localization which, however, may be weaker than exponential \cite{fleishman1977fluctuations,inui1994unusual,cheraghchi2005localization}. 
In the presence of long-range exchange interactions, the (single-excitation) localization properties of the system are more subtle \cite{deng2018duality,mirlin1996transition,nosov2019robustness,de2005localization,nosov2019,kutlin2020} and many features still merit further investigation, which is one of the motivations of the present work. 

Specifically, we study long-range interacting disordered spin chains -- collection of two-level atoms, molecules or spins arranged in nearly periodic quasi one-dimensional array and coupled with each other by the resonant dipole-dipole exchange interaction.
We raise the questions whether or not, and to what degree, such a disordered system can serve for excitation or spin polarization transfer between the sender and the receiver spins coupled to the opposite ends of the chain in a controllable way.
To that end, we first determine the localization properties of the system and their dependence on the energy, comparing and contrasting the long-range and nearest-neighbor interacting spin systems. Obviously, only chains of length smaller or comparable to the longest localization length can transfer excitation between the two ends.  
Next we identify the energy eigenstates of the chain that have sufficient support at the two ends of the chain to strongly couple to the sender and receiver spins. 
We then explore two excitation transfer protocols, one that involves static resonant couplings of the sender and receiver spins to the most suitable eigenstate of the chain \cite{yao2011robust,zwick2014optimized,zwick2014optimized},
and the other inspired by stimulated Raman adiabatic transfer \cite{kuklinski1989adiabatic,STIRAP-RMP1998,vitanov2017stimulated} that involves counterintuitive time-dependent couplings of the sender and receiver spins to the corresponding eigenstate of the chain. We find that the adiabatic coupling, despite being slower than the static coupling scheme, leads to a much higher probability of excitation transfer as it is more robust to various sources of disorder.  

The paper is organized as follows. 
in Sec. \ref{sec:system} we introduce the Hamiltonian of the system involving a collections of spins (two-level systems) with long-range resonant dipole-dipole interactions. 
In Sec. \ref{sec:loc} we consider disordered spin chains and numerically determine the localization lengths for different single-excitation eigenstates of the system
in the presence of energy (diagonal) and position (off-diagonal) disorder. 
In Sec. \ref{sec:prot} we present two excitation transfer protocols between the sender and receiver spins resonantly coupled to a suitable energy eigenstate 
of the disordered spin chain.
In Sec. \ref{sec:mean} we extract the mean transfer probability for chains of different length with different strength and type of disorder. 
Our conclusions are summarized in Sec. \ref{sec:conc}.

\section{The system} 
\label{sec:system}

\begin{figure}[t]
\includegraphics[width=0.9\columnwidth]{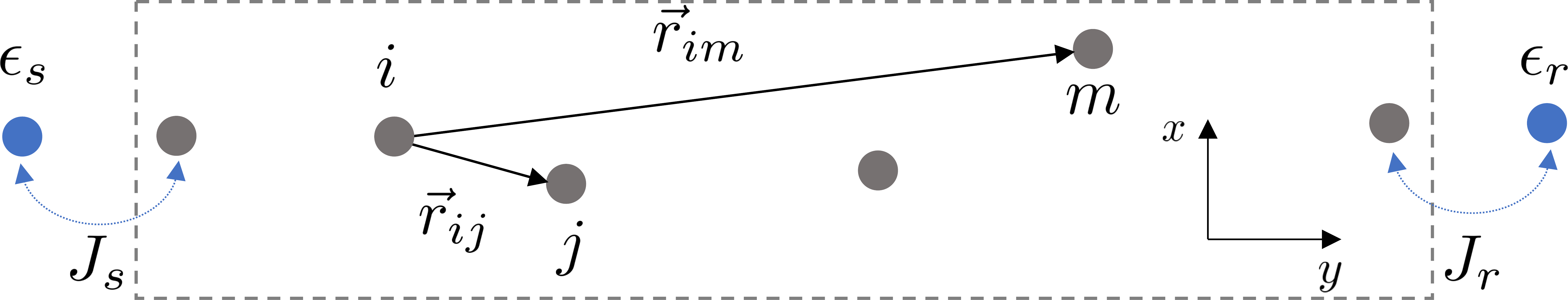}
\caption{Schematic of a position (and energy) disordered chain of spins $i,j, \ldots , m, \ldots$ in the $xy$ plane. 
The spin chain is coupled with rates $J_{s,r}$ to the sender (s) and receiver (r) spins having energies $\epsilon_{s,r}$.}
\label{fig:scheme}
\end{figure}

We consider a chain of $N$ spins -- two-level systems -- interacting with each other via the long-range exchange interactions $J_{ij} = C_{3} (1-3\cos^{2} \theta_{ij})/\abs{\vec{r}_{ij}}^{3}$, where $C_{3} \propto |\vec{\wp}|^2$  
is the electric or magnetic dipole-dipole interaction coefficient, 
$\vec{r}_{ij}$ is the position vector between spins $i$ and $j$, and 
$\theta_{ij}$ is the angle between the direction of the dipole moments $\vec{\wp}$ and the position vector between the spins. 
We account only for the near-field part of the total dipole-dipole interaction potential and neglect the retardation and spontaneous radiative decay of the spin excitations \cite{Lehmberg1970,Craig1984}, assuming that the typical distance between the spins is much smaller than the wavelength of the transition between the spin-up and spin-down states. 
The Hamiltonian of the system is
\begin{equation}\label{eq:Hamdd}
\mathcal{H} = \frac{1}{2} \sum_{i=1}^{N} \epsilon_{i} \hat{\sigma}_{i}^{z} 
+ \sum_{i \neq j}^N J_{ij} ( \hat{\sigma}^{+}_{i} \hat{\sigma}^{-}_{j}
+ \hat{\sigma}^{+}_{j} \hat{\sigma}^{-}_{i}) , 
\end{equation}
where $\epsilon_{i}$ is the excitation energy of spin $i$, 
$\hat{\sigma}^{x,y,z}_{i}$ are the Pauli spin operators
and $\hat{\sigma}^{\pm}_i = \frac{1}{2} (\hat{\sigma}^x_i \pm i \hat{\sigma}^y_i)$ are the raising and lowering operators.
We assume that all the spins are positioned in one ($xy$) plane (see Fig.~\ref{fig:scheme}) and their dipole moments ($\vec{\wp} \parallel \hat{z}$) are perpendicular to that plane, $\theta_{ij}=\pi/2 \; \forall \; i,j$, thus $J_{ij} = C_{3}/\abs{\vec{r}_{ij}}^{3}$. 

We assume that a sender and a receiver spins are coupled in controllable way to the opposite ends of a finite spin chain, see Fig.~\ref{fig:scheme}.
In order to transfer an excitation between the sender and a receiver spins, the disordered chain should possess extended eigenstates having support at its two ends. 
To selectively couple the sender and receiver spins to the suitable eigenstates of the chain, we assume that their
energies $\epsilon_{s}$, $\epsilon_{r}$ and couplings $J_{s}$, $J_{r}$ to the first and last spins of the chain can be precisely controlled, unlike the energies and couplings of the spins in the disordered chain. 
Initially, the excitation is localized at the sender spin, while the spin chain contains no excitations, and our aim will be to retrieve the excitation from the receiver spin at a specific time $\tau$ to be determined below. 

We next examine the localization length of the single-excitation eigenstates of spin chains in the presence of diagonal disorder corresponding to energy disorder of individual spins, and off-diagonal disorder in the interspin couplings stemming from the position disorder of the spins.

\section{Localization lengths in disordered spin chains} 
\label{sec:loc}

We impose diagonal disorder corresponding to random variations of the spin excitation energies 
$\epsilon_{j} = \epsilon_0 + \delta \epsilon_j$ around some $\epsilon_0$ (which can be set to 0) 
with $\delta \epsilon_j$ having a Gaussian probability distribution 
$P(\delta \epsilon)=\frac{1}{\sqrt{2 \pi \sigma_{\epsilon}^2}}e^{-\frac{\delta \epsilon^2 }{2\sigma_{\epsilon}^2}}$ 
with the mean $\langle \delta \epsilon \rangle= 0$ and variance $\sigma_{\epsilon}^2$. 
Next, the position of each spin $j$ is given by the coordinates $(x_{j},y_{j})$. 
In an ideal 1D lattice with period $a$, we would have $x_j = aj$ and $y_j=0$ for all spins $j =1,2,\ldots, N$, 
and the exchange interaction strength between the nearest-neighbor spins would be $J=C_3/a^3$, 
the next-nearest neighbors $J/2^3$, etc. 
We impose the position disorder via $x_{j} \to aj + \delta x_{j}$ and $y_{j} \to \delta y_{j}$, where the random variables $\delta x_{j}$ and $\delta y_{j}$ have a Gaussian probability distribution $P(\delta \mu)=\frac{1}{\sqrt{2 \pi \sigma_{\mu}^2}}e^{-\frac{\delta \mu^2}{2\sigma_{\mu}^2}}$ ($\mu=x,y$) around mean $\langle \delta \mu \rangle=0$ with variance $\sigma_{\mu}^2$.
The position disorder then translates to off-diagonal (interspin coupling) disorder in the Hamiltonian (\ref{eq:Hamdd}). 

In the limit of $N \to \infty$, disorder leads to (Anderson) localization of all the eigenstates of the system 
\cite{mott1961theory,anderson1958absence,gogolin1982}.
The wavefunction $\psi_k(x)$ of each single-excitation eigenstate $\ket{\psi_k}$ is then localized around some position $\mu_k$ with the localization length $\xi_{k}$.
An important characteristic of the system is the dependence of the localization length $\xi_{k}$ on the energy $E_k$ of the eigenstates to be used for the excitation transfer. 
To determine the localization length, we numerically diagonalize the Hamiltonian for sufficiently long chains ($N=1000$ spins) to neglect the finite size effects, 
and then for each eigenstate we identify the position $\mu_k$ corresponding to the maximum (in absolute value) of the wavefunction $\psi_k(x)$ and subsequently fit an exponential function 
\begin{equation}
  |\psi_k(x)| \propto e^{-\frac{|x-\mu_k|}{\xi_{k}}}  \label{eq:psiloc}
\end{equation}
to the spatial profile of the eigenstate, extracting thereby the localization length $\xi_{k}$. 
We note that the thus obtained localization length is a convenient measure of the spatial extent of the wavefunction even if it is not exponentially localized (see below). 

A more common measure to quantify the localization properties of the eigenstates is the inverse participation ratio (IPR) \cite{edwards1972numerical}. 
It is, however, not suitable for our purposes, since IPR cannot determine whether a wavefunction is spatially localized on a number of neighboring sites or is delocalized on a similar number of remote sites \cite{IPRdef}. 
We use, therefore, an alternative method to verify that the localization length $\xi_k$ extracted from the exponential fit (\ref{eq:psiloc}) is a reliable quantity to characterize our system. 
We can partition the chain into two halves and for each eigenstate $\ket{\psi_k} = \sum_{i=1}^N v_i^{(k)} \ket{i}$ calculate the excitation number variance in one of the halves \cite{kiefer2022}, 
\begin{equation}
\Delta n^2_k= \ev{\hat{n}^2} - \ev{\hat{n}}^2,
\end{equation}
where $\hat{n} = \sum_{i=1}^{N/2} \hat{\sigma}^{+}_{i} \hat{\sigma}_{i}^{-}$ is the excitation number operator with eigenvalues $n=0,1$ since we consider only single-excitation states. The variance is therefore given by 
\begin{equation}
\Delta n^2_k= p_k-p_k^2 ,
\end{equation}
where $p_k = \sum_{i=1}^{N/2} |v_i^{(k)}|^2$ is the probability to find the excitation in the left half of the chain. 

\begin{figure*}
\includegraphics[width=\textwidth]{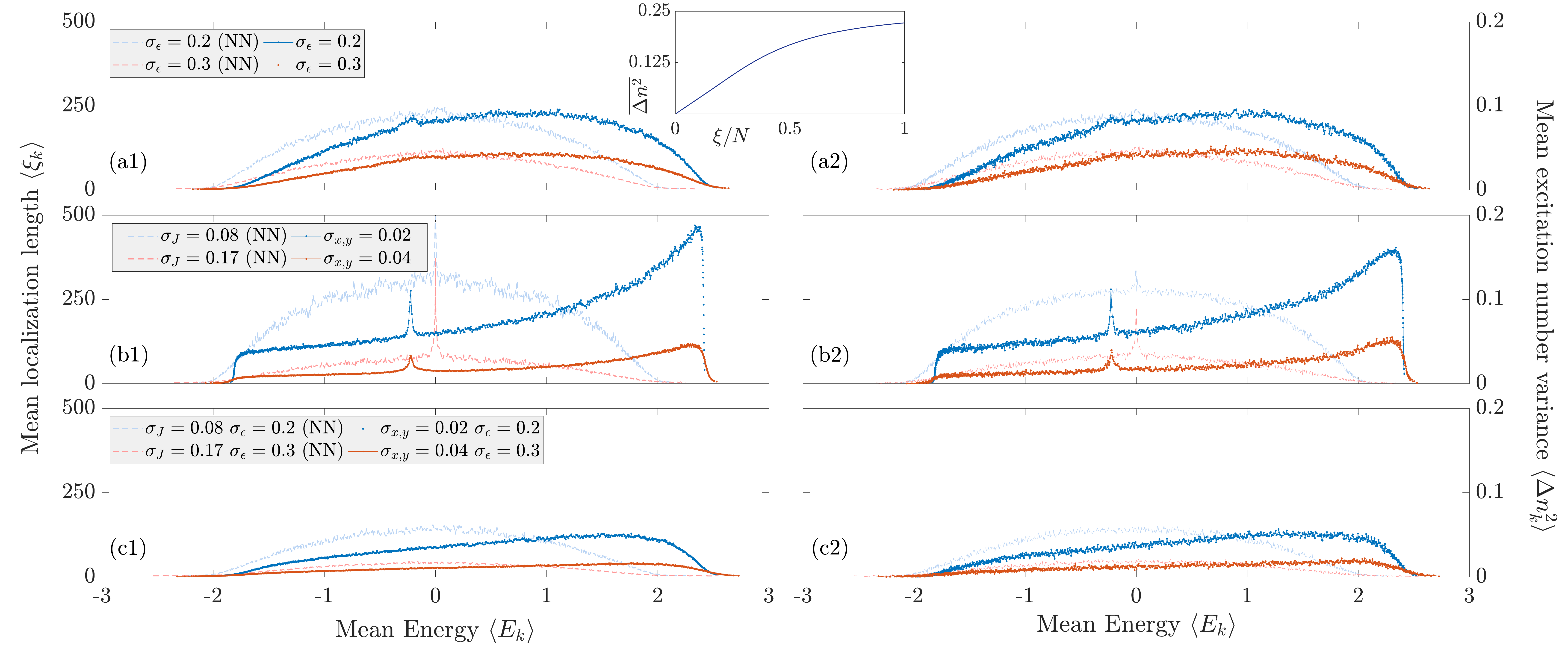}
\caption{Mean localization length $\langle \xi_{k} \rangle$ (in units of lattice spacing $a=1$) [left panels (a1), (b1), (c1)], and mean excitation number variance $\langle \Delta n^2_k \rangle$ [right panels (a2), (b2), (c2)] 
vs the mean energy $\langle E_{k} \rangle$ (in units of $J = C_3/a^3$) of the $k$-th eigenstate of a chain of $N=1000$ spins obtained upon averaging over $1000$ independent realizations of disordered chains with long-range interactions (solid lines with filled circles) and nearest-neighbor interactions (dashed lines), for   
(a) energy (diagonal) disorder with standard deviation $\sigma_{\epsilon}$, 
(b) position (off-diagonal) disorder with standard deviation $\sigma_{x,y}$ or $\sigma_{J}$, and 
(c) combination of energy and position disorder.
For illustrative purposes, we use in (a) and (b) the strength of the diagonal $\sigma_{\epsilon}$ and  
off-diagonal $\sigma_{x,y}$ (or $\sigma_{J}$) disorders that lead to comparable localization lengths.
Inset shows the averaged number variance $\overline{\Delta n^2}$ vs $\xi/N$, as described in the text.}
\label{fig:sn}
\end{figure*}

Clearly, for a strongly localized state with $\xi/a \ll N/2$, the probability $p$ is either close to 0 or close to 1 (unless the wavefunction is localized near the center of the chain, $\mu/a \simeq N/2$, the probability of which is $2\xi /(aN) \ll 1$), and the number variance is small, $\Delta n^2 \to 0$. 
In the opposite limit of a completely delocalized wavefunction $\xi/a > N$, the probability is $p \simeq 1/2$ and the number variance approaches the maximum $\Delta n^2 \to 1/4$. 
Assuming an exponentially localized  wavefunction $\psi(x)$ of the form (\ref{eq:psiloc}), we can calculate $p$ for any position of the peak $\mu$, and upon averaging over the peak positions $\mu/a \in [1,N]$ we obtain a relation between $\overline{\Delta n^2}$ and $\xi/N$ shown in the inset of Fig.~\ref{fig:sn}. 
For small $\xi/a < N/2$, the number variance grows approximately linearly with the localization length as $\overline{\Delta n^2} \approx \frac{3}{8} \frac{\xi}{aN}$, and it starts to saturate thereafter. 

In Fig.~\ref{fig:sn} (left panels: a1, b1, c1), we show the mean localization length $\langle \xi_{k} \rangle$ versus 
the mean energy $\langle E_{k} \rangle$ of the eigenstate for three different cases: 
(a) diagonal (energy) disorder, 
(b) off-diagonal (position) disorder, and 
(c) combination of diagonal and off-diagonal disorders. 
The corresponding mean excitation number variances $\langle \Delta n_k^2 \rangle$ are shown in Fig.~\ref{fig:sn} (right panels: a2, b2, c2). 
For each case we consider two different strengths of the disorder determined by the standard deviations $\sigma_{\epsilon}$ and $\sigma_{x,y}$. 

For comparison, we also consider chains with nearest-neighbor interactions and the same effective disorder as described by Hamiltonian
\begin{equation}\label{eq:HamNN}
\mathcal{H}_{\mathrm{nn}}= \frac{1}{2} \sum_{i=1}^{N} \epsilon_{i} \hat{\sigma}_{i}^{z}
+ \sum_{i=1}^{N-1} J_i (\hat{\sigma}^{+}_{i} \hat{\sigma}_{i+1}^{-} + \hat{\sigma}^{+}_{i+1} \hat{\sigma}_{i}^{-}) , 
\end{equation}
where $\epsilon_i$ are the random spin energies as above, while $J_i = J + \delta J_i$ are the exchange couplings 
with $J=C_3/a^3$ and $\delta J_i$ being Gaussian random variables with the mean $\langle \delta J \rangle= 0$ and 
standard deviation determined by the error propagation formula
\[
\sigma_{J} \approx \abs{ \partial_x D(x,y)} \sigma_x + \abs{\partial_y D(x,y)} \sigma_y, 
\] 
where $D(x,y) = C_3/(x^2 + y^2)^{3/2}$.

Note that, in an ideal lattice with no disorder, the single excitation spectrum of Hamiltonian~(\ref{eq:Hamdd}) is given by   
\begin{equation} \label{eq:lrs}
E_{k} = 2 \sum_{m=1}^{N} \frac{J}{m^3}\cos{\frac{\pi k m}{N+1}} ,
\end{equation}
while the spectrum of the system with only the nearest-neighbor interactions, Eq.~(\ref{eq:HamNN}), 
corresponds to the $m=1$ term in the above sum, i.e. $E_{k}^{(\mathrm{nn})}=2 J \cos{\frac{\pi k} {N+1}} \; \in [-2J,2J]$. 
One can treat perturbatively the $m>1$ terms of Eq. (\ref{eq:lrs}) near the band edges and deduce \cite{malyshev1995hidden,kozlov1998zero} that the lower edge of the energy band is shifted from $-2J$ to approximately $-1.8J$ while upper edge is shifted from $2J$ to approximately $2.4J$. Thus, the long-range character of the interaction affects the energy band structure and the density of states. 

\paragraph{Diagonal disorder.} 
Consistent with the above discussion, for a chain with long-range interactions and diagonal disorder, we observe in Fig.~\ref{fig:sn}(a1) and (a2) that the profile of the mean localization length $\langle \xi_{k} \rangle$ and the nearly identical profile of the mean excitation number variance $\langle \Delta n_k^2 \rangle$ are shifted and skewed towards the higher energies $\langle E_{k} \rangle$, as compared to the nearest-neighbor interacting chains.
For the presently considered dipole-dipole interactions, $J_{ij} \propto 1/|r_{ij}|^3$, the localization length $\langle \xi_{k} \rangle$ remains finite for all energies $\langle E_{k} \rangle$. 
We note, however, that for power-law interaction $J_{ij} \propto 1/|r_{ij}|^\nu$ with decreasing $\nu$ a localization-delocalization transition occurs at $\nu=3/2$ near the (shifted) upper edge of the energy band $\langle E_{k} \rangle \approx 5J$ \cite{rodriguez2003anderson}.

\paragraph{Off-diagonal disorder.} 
Even though the wavefunctions of the eigenstates of a chain with off-diagonal disorder may not be exponentially localized for all energies, for consistency and comparison with diagonal disorder, we still use the exponential fit of Eq.~(\ref{eq:psiloc}) to deduce the localization length and verify its applicability by the corresponding excitation number variance.
For the nearest-neighbor interacting chain with only off-diagonal disorder, the first feature to note in Fig.~\ref{fig:sn}(b1, b2) 
is the sharp peak of the localization length at zero energy. 
This peak is related to the well-known divergence of the density of states $\rho(E) \sim \frac{1}{E \abs{\ln{E}}^3}$ \cite{theodorou1976extended,eggarter1978singular} leading to the localization length divergence as $\xi \sim \abs{\ln{E}}$ that follows from the Thouless relation \cite{thouless1972relation}. 
But unlike the case of diagonal disorder, the eigenstates near zero energy are localized as 
$|\psi (x)| \propto e^{-\sqrt{x/\zeta}}$ rather than exponentially \cite{fleishman1977fluctuations,inui1994unusual,cheraghchi2005localization}. 
We note the relevant early studies of Dyson \cite{DysonComm,dyson1953dynamics} and the insightful connection to the graph theoretical concepts \cite{bipartiteComm,inui1994unusual}.

The long-range interactions in the chain with off-diagonal disorder \cite{levitov1990delocalization,kozlov1998zero,klinger2021single} 
lead to certain modification of the localization spectrum. 
The zero-energy peak of the nearest-neighbor interacting chain is now displaced to $\langle E_{k} \rangle \simeq -0.22J$, which follows from the perturbative treatment of Eq. (\ref{eq:lrs}) near the center of the band \cite{kozlov1998zero}, and is suppressed, 
since the underlying lattice is weakly non-bipartite due to the weak next-nearest-neighbor interactions \cite{bipartiteComm}, which is in complete agreement with our numerical results in Fig. \ref{fig:sn}(b1, b2). 
We note again that the use of IPR \cite{IPRdef} is inadequate to quantify the localization length in the vicinity of $\langle E_{k} \rangle \simeq -0.22J$, as it would indicate more, rather than less, localized states \cite{kozlov1998zero}. That is why we still use the localization length $\langle \xi_k \rangle$ obtained from the exponential fit of Eq.~(\ref{eq:psiloc}) and verify its applicability by the corresponding excitation number variance $\langle \Delta n^2 \rangle$. 

Another feature is that, perhaps counterintuitively, disordered chains with long-range interactions exhibit shorter localization length in the central part of the spectrum, 
as compared to chains with only nearest-neighbor interactions \cite{deng2018duality,nosov2019robustness,de2005localization}; in effect the long-range interactions amplify the disorder. 
But for larger energies the localization length $\langle \xi_{k} \rangle$ (and the excitation number variance $\langle \Delta n^2_k \rangle$)
gradually increases \cite{kozlov1998zero,fidder1991optical} and it exhibits a sharp peak near the upper edge of the energy band, $\langle E_{k} \rangle \approx 2.4J$. The states near the upper edge of the energy band are in fact completely delocalized, $\langle \xi_{k} \rangle \approx N/2$, at least for not too strong off-diagonal disorders that we consider. This behaviour is reminiscent to the emergence of extended states at the band edge for spin chains with diagonal disorder and long-range interactions $J_{ij} \propto 1/|r_{ij}|^\nu$ with decreasing power $\nu$, but for our case of off-diagonal disorder and $\nu=3$, the sharp peak is much more pronounced.

\paragraph{Combined diagonal and off-diagonal disorder.}
Finally in Fig.~\ref{fig:sn}(c1, c2) we show the mean localization length and the mean excitation number variance versus the mean energy 
for the chains with both diagonal and off-diagonal disorders that concurrently localize the system eigenstates. 
Now the (shifted) zero-energy peak is completely suppressed \cite{bipartiteComm} while the eigenstates with the longest localization length reside between the center and the upper edge of the band skewed by the long-range interactions. 

To summarize, the important information gained by our analysis of the localization lengths in disordered spin chains is the maximum length of a finite chain that can support excitation transfer through an extended eigenstate. 
Conversely, when the chain length exceeds the maximum localization length of the eigenstates, we expect the transfer to be completely suppressed.
We note that in all cases when the obtained mean localization length is sufficiently shorter than the chain length, $\langle \xi_k \rangle < aN/2$, the relation $\langle \Delta n^2_k \rangle \approx \frac{3}{8} \frac{\langle \xi_k \rangle}{aN}$ holds to a very good approximation,
which justifies our approach to characterizing the localization properties of disordered, long-range interacting spin chains.

\section{Excitation Transfer schemes} 
\label{sec:prot}

The large localization length in a disordered spin chain is necessary but not yet sufficient 
to ensure efficient transfer of excitation between the sender and receiver spins. 
Rather, the extended eigenstates of the chain should have sufficient support at the two ends 
of the chain in order to strongly couple to the sender and receiver spins. 

Consider again the spin chain with long-range interactions and no disorder. Solving the eigenvalue problem
\begin{equation}
\mathcal{H} \ket{\psi_{k}} = E_{k} \ket{\psi_{k}} ,
\end{equation}
we obtain the eigenstates $\ket{\psi_{k}}= \sum_i v_{i}^{(k)} \ket{i}$
which couple to the sender and receiver spins at the two ends of the chain with the corresponding strengths 
\begin{equation}
\Omega_{s}^{(k)}= J_{s} v_{1}^{(k)}, \quad \Omega_{r}^{(k)}= J_{r} v_{N}^{(k)}, \label{eq:coupls}
\end{equation}
where $J_{s}$ and $J_{r}$ are the coupling strength of the sender and receiver spins to the first and the last spins of the chain. 
Hence, in order to efficiently transfer the excitation from the sender to the receiver spin via a particular eigenstate $\ket{\psi_{k}}$ of the
chain, this eigenstate should have large amplitudes $|v_{1,N}^{(k)}|$ at both ends of the chain.

\begin{figure}[t]
\includegraphics[width=0.8\columnwidth]{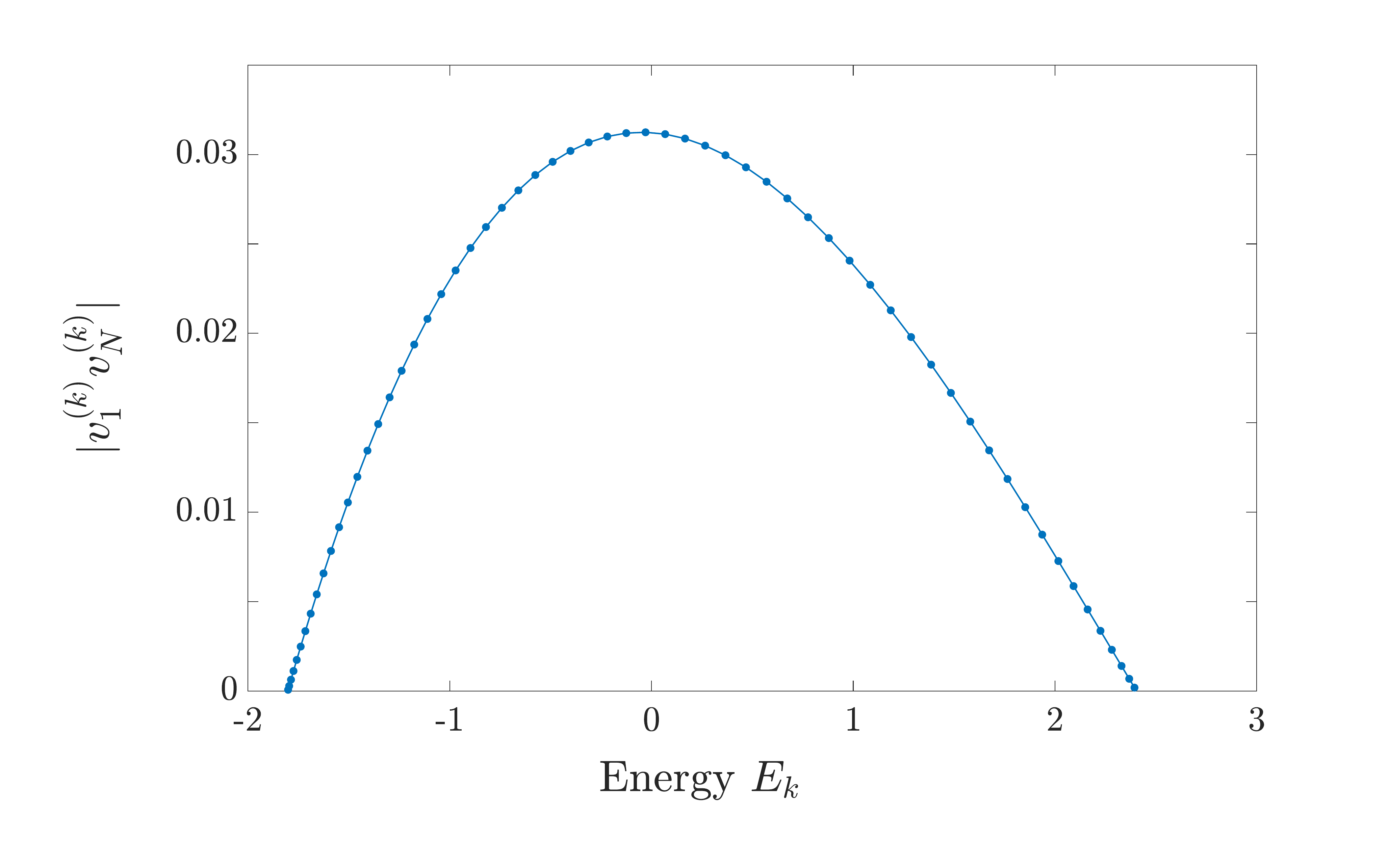}
\caption{Absolute value of the product $| v_{1}^{(k)} v_{N}^{(k)}|$ of the boundary amplitudes of $k$-th eigenvector  
of the chain vs the eigenenergy $E_k$ (in units of $J$), for a chain of $N=61$ spins with no disorder.}
\label{fig:support}
\end{figure}

In Fig. \ref{fig:support} we show the absolute value of the product $|v_{1}^{(k)} v_{N}^{(k)}|$ of the boundary amplitudes of the different energy eigenstates $\ket{\psi_{k}}$ of the chain with no disorder. 
This figure reveals that the eigenstates most suitable for the transfer are in the middle of the spectrum, $E_k \sim 0$, while the eigenstates at the upper edge of the spectrum, $E_k \lesssim 2.4 J$, would only weakly couple to the sender and receiver spins and are thus unsuitable for the excitation transfer, despite having large (or even divergent) localization length in disordered chains. 
Having in mind the chains with both diagonal and off-diagonal disorder exhibiting the localization peak in the vicinity of $E = -0.22 J$, we shall tune the energies of the sender and receiver spins to $\epsilon_{s,r} \approx -0.22J$.

Another critical issue for the efficient transfer via the selected eigestates of the chain is the small leakage of excitation, initially at the sender spin, to all other non-resonant eigenstates of the chain \cite{yao2011robust,zwick2014optimized}. 
In a chain of $N$ spins, the average distance between the energy eigenstates is $\Delta E \simeq 4J/N$. Therefore, the coupling strength of the sender and receiver spins, tuned to resonance to a particular eigenstate, should satisfy $\Omega_{s,r} < \Delta E$. 
Since the amplitudes of the edge states for the most delocalized eigenstates are $v_{1,N}^{(k)} \sim 1/\sqrt{N}$, 
we obtain from (\ref{eq:coupls}) that the coupling rates should satisfy $J_{s,r} \lesssim J/\sqrt{N}$ in order to avoid the leakage of the excitation to the undesired states of the chain and attain high transfer probability \cite{wojcik2005unmodulated}.

\begin{figure}[t]
\includegraphics[width=1.0\columnwidth]{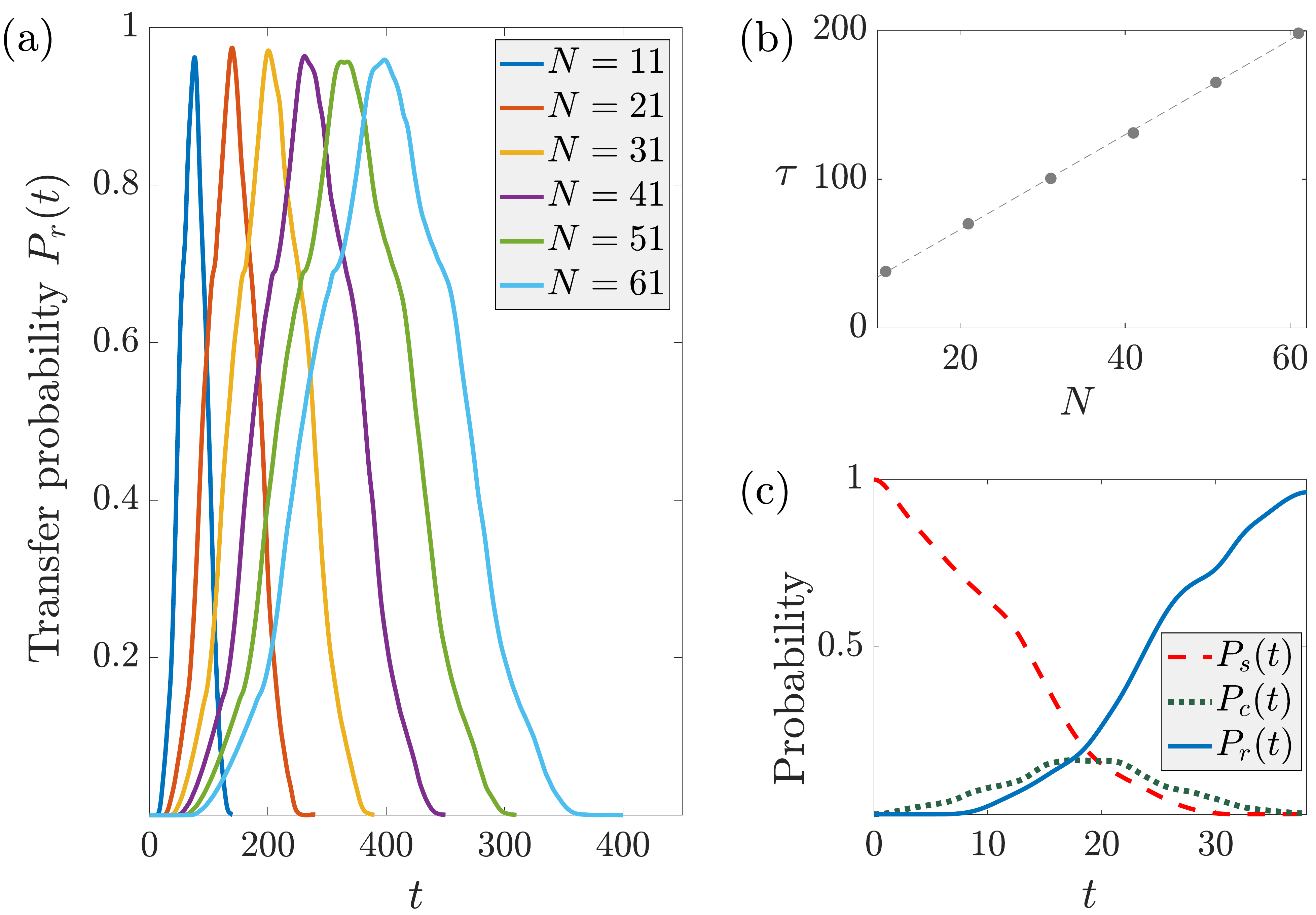}
\caption{Excitation transfer via static couplings of the sender and receiver spins with rates $J_{s,r}=0.49 J/ \sqrt{N}$ to the chain of $N$ spins with no disorder.
(a)~Transfer probability $P_{r}(t)$ vs time $t$ (in units of $1/J$) for different chain lengths $N$. 
The energies of the sender and receiver spins $\epsilon_{s,r}$ are tuned to the energy of the eigenstate of the chain closest to $E = -0.22J$. 
(b)~Transfer time $\tau$ (gray filled circles), corresponding to the first peak of the transfer probability in (a) for each chain length $N$.
Dashed line shows the linear fit $\tau J = 3.2 N + 2.3$. 
(c)~Time-evolution of the excitation probability for the sender $P_{s}(t)$, receiver $P_{r}(t)$ and intermediate chain $P_{c}(t)$, for a chain of $N=11$ spins.}
\label{fig:sttrO}
\end{figure}

\paragraph{Static coupling to the chain.}
To illustrate the ongoing discussion, in Fig.~\ref{fig:sttrO} we show the dynamics of excitation transfer 
between the sender and receiver spins via spin chains of different length $N$ with no disorder. 
For convenience, we chose chains with odd number of spins, $N=11,21,\ldots$, and tune the energies of the sender and receiver spins $\epsilon_{s,r}$ to the energy of the ``fittest'' eigenstate closest to $E = -0.22J$. 

The state of the system in the single excitation subspace can be written as $\ket{\Psi} = \alpha_s \ket{s} + \sum_{i=1}^N \alpha_i \ket{i} + \alpha_r \ket{r}$, where $\alpha_j$ are the amplitudes and $\ket{j}$ denotes the state with the excitation at position $j=s,r$ or $i \in [1,N]$. 
Initially the excitation is localized at the sender spin, $\ket{\Psi(0)} = \ket{s}$, and the couplings $J_{s,r}$ are set to the constant values $J_{s,r} \simeq 0.5J/\sqrt{N}$. 
The state of the system $\ket{\Psi (t)}$ evolves according to the Hamiltonian (\ref{eq:Hamdd}), and the transfer probability to the receiver spin $P_r(t) = \abs{\bra{r}\ket{\Psi(t)}}^2$ is shown in Fig.~\ref{fig:sttrO}(a). 
In a three-state system, complete transfer would occur at time $\tau = \pi/(2\sqrt{2} \Omega_{s,r})$. 
Our multilevel system now behaves as an effective three-state system with a single intermediate eigenstate of the chain, and the transfer time scales as $\tau \propto N$ consistently with $\Omega_{s,r} \propto 1/N$, see Fig.~\ref{fig:sttrO}(b). In Fig.~\ref{fig:sttrO}(c) we show the dynamics of probabilities of excitation of the sender spin, $P_s(t) = \abs{\bra{s}\ket{\Psi(t)}}^2$, the chain, $P_c(t) = \sum_{i=1}^N \abs{\bra{i}\ket{\Psi(t)}}^2$, and the receiver spin, $P_r(t)$, during one full transfer cycle.

\paragraph{Time-dependent adiabatic couplings.}
In a three-state system, a more efficient excitation transfer can be achieved using an analog of stimulated Raman adiabatic passage (STIRAP) \cite{STIRAP-RMP1998,vitanov2017stimulated,kuklinski1989adiabatic}.
It involves time-dependent couplings and must be sufficiently slow in order to be adiabatic, but is robust and avoids populating the intermediate -- here the spin-chain -- state(s).

Consider an effective three-state system $\ket{\Psi} = \alpha_s \ket{s} + \alpha_k \ket{\psi_k} + \alpha_r \ket{r}$ governed by the Hamiltonian 
\begin{equation}
\mathcal{H}^{\mathrm{eff}} = \Delta \epsilon_k \ket{\psi_k}\bra{\psi_k} + (\Omega_s^{(k)} \ket{s}\bra{\psi_k} + \Omega_r^{(k)} \ket{r}\bra{\psi_k} + \mathrm{H. c.})
\end{equation}
where $\Delta \epsilon_k = E_k - \epsilon_{s,r}$ is a possible energy mismatch between the selected eigenstate of the chain $\ket{\psi_k}$ and the sender and receiver spins.  
This Hamiltonian has a zero-energy coherent population trapping (or dark) eigenstate 
$\ket{\Psi_0} \propto \Omega_r^{(k)} \ket{s} - \Omega_s^{(k)} \ket{r}$ that does not involve the intermediate state $\ket{\psi_k}$ of the spin chain. 
With the excitation initially localized on the sender spin, we set the coupling $|\Omega_r^{(k)}| \gg |\Omega_s^{(k)}|$ such that the dark state 
coincides with the initial state, $\ket{\Psi_0} = \ket{s}$. We then slowly switch off $\Omega_r^{(k)}$ and switch on $\Omega_s^{(k)}$, which results 
in an adiabatic rotation of the dark state $\ket{\Psi_0}$ towards $\ket{r}$, and at the final time $\tau$, when $|\Omega_r^{(k)}| \ll |\Omega_s^{(k)}|$, we obtain
$\ket{\Psi_0} \simeq \ket{r}$. To realize this so-called counterintuitive pulse sequence, we use the time-dependent boundary couplings
 \begin{equation}
J_{s,r}(t)= \frac{J_{s,r}^{\mathrm{max}}}{2} \Big(1 \pm  \tanh{(\gamma t/\tau - \beta_{s,r})}\Big) , \label{eq:stirapulses}
\end{equation}
where $J_{s,r}^{\mathrm{max}} \simeq 0.5/\sqrt{N}$ as before, while the parameters $\gamma = 6$, $\beta_{s,r} = 2.3,3.6$ and the process duration $\tau \propto N$
are chosen so as to optimize the overlap between the pulses and achieve adiabaticity with sufficiently large effective pulse area
$\int_0^{\tau} dt \sqrt{|\Omega_s^{(k)}(t)|^2 + |\Omega_r^{(k)}(t)|^2} \gtrsim 10$ \cite{STIRAP-RMP1998,vitanov2017stimulated}. 
We note that the adiabatic population transfer has been applied to multilevel systems before \cite{Shore1991,vitanov1999adiabatic}.

\begin{figure}[t]
\includegraphics[width=1.0\columnwidth]{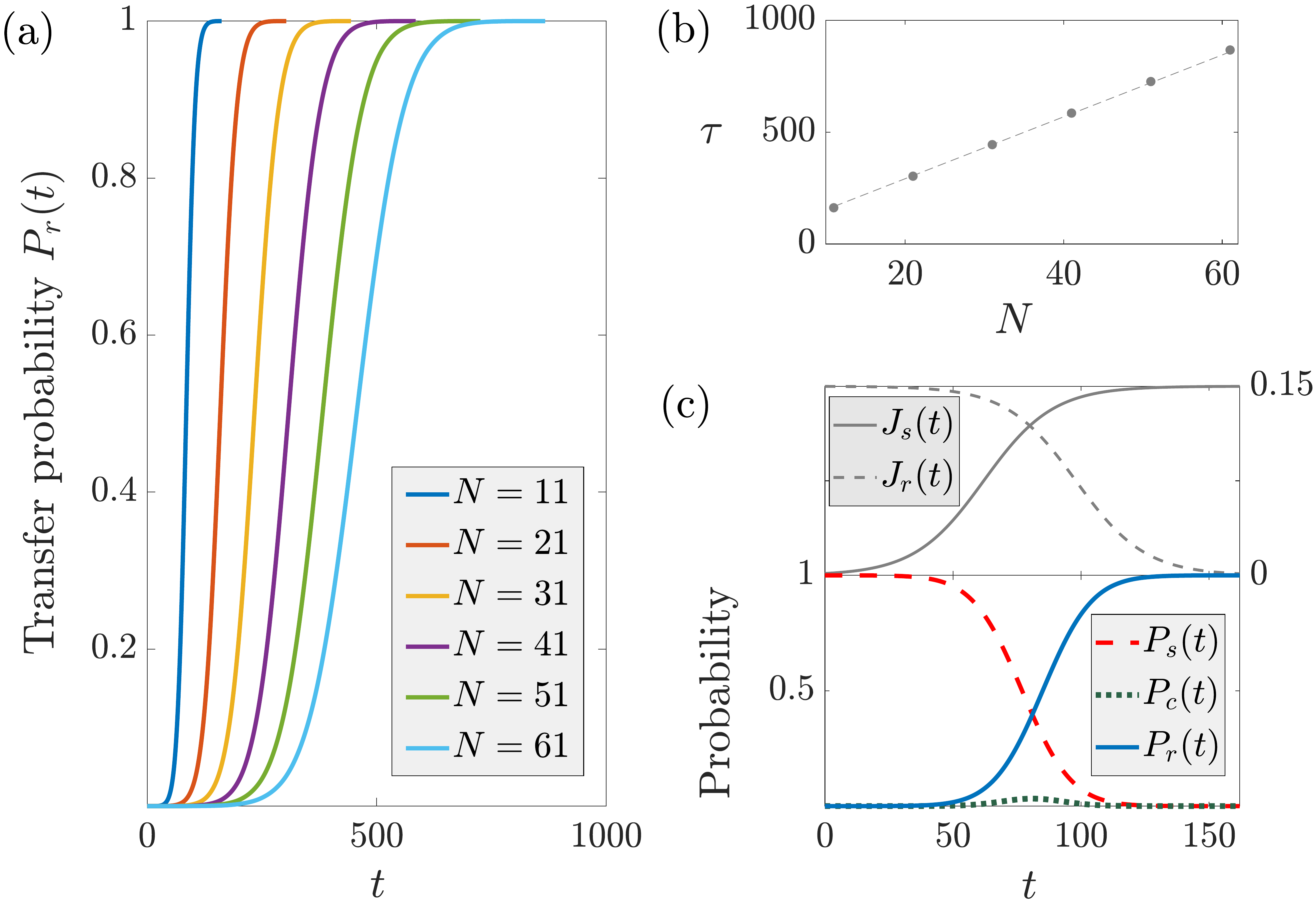}
\caption{Stimulated adiabatic transfer of excitation between the sender and receiver spin using time-dependent couplings of Eq.~(\ref{eq:stirapulses}), for a chain with no disorder. 
(a)~Transfer probability $P_{r}(t)$  vs time $t$ (in units of $1/J$) for chains of different length. 
(b)~Transfer time $\tau$ (gray filled circles) as a function of $N$, and the linear fit $\tau J = 14.1N + 6.9$ (dashed line). 
(c)~Top panel shows the time-dependent coupling rates of rates $J_{s,r}(t)$ of Eq.~(\ref{eq:stirapulses}), and 
the bottom panel shows the dynamics of excitation probabilities of the sender $P_{s}(t)$, receiver $P_{r}(t)$ and intermediate chain $P_{c}(t)$, for $N=11$.}
\label{fig:stirapO}
\end{figure}

In Fig.~\ref{fig:stirapO} we illustrate the adiabatic transfer protocol for chains of different length and time-dependent couplings of Eq.~(\ref{eq:stirapulses}) but otherwise the same parameters as in Fig.~\ref{fig:sttrO}. 
We achieve nearly perfect population transfer for all considered cases, see Fig.~\ref{fig:stirapO}(a), at the expense of longer duration of the process $\tau$, see Fig.~\ref{fig:stirapO}(b). 
Note that during the transfer, as the system adiabatically follows the coherent population trapping state $\ket{\Psi_0}$, the chain contains almost no excitation at all times, Fig.~\ref{fig:stirapO}(c).

\section{Transfer probability in disordered chains} 
\label{sec:mean}

Having determined the localization lengths $\xi$ in long disordered spin chains in Sec.~\ref{sec:loc} and 
potentially suitable excitation transfer protocols in Sec.~\ref{sec:prot}, we now analyze the mean probability $\langle P_{r} \rangle$ 
of excitation transfer between the sender and receiver spins via disordered spin chains of finite length $N$ comparable to $\xi$. 

\begin{figure}[t]
\includegraphics[width=0.9\columnwidth]{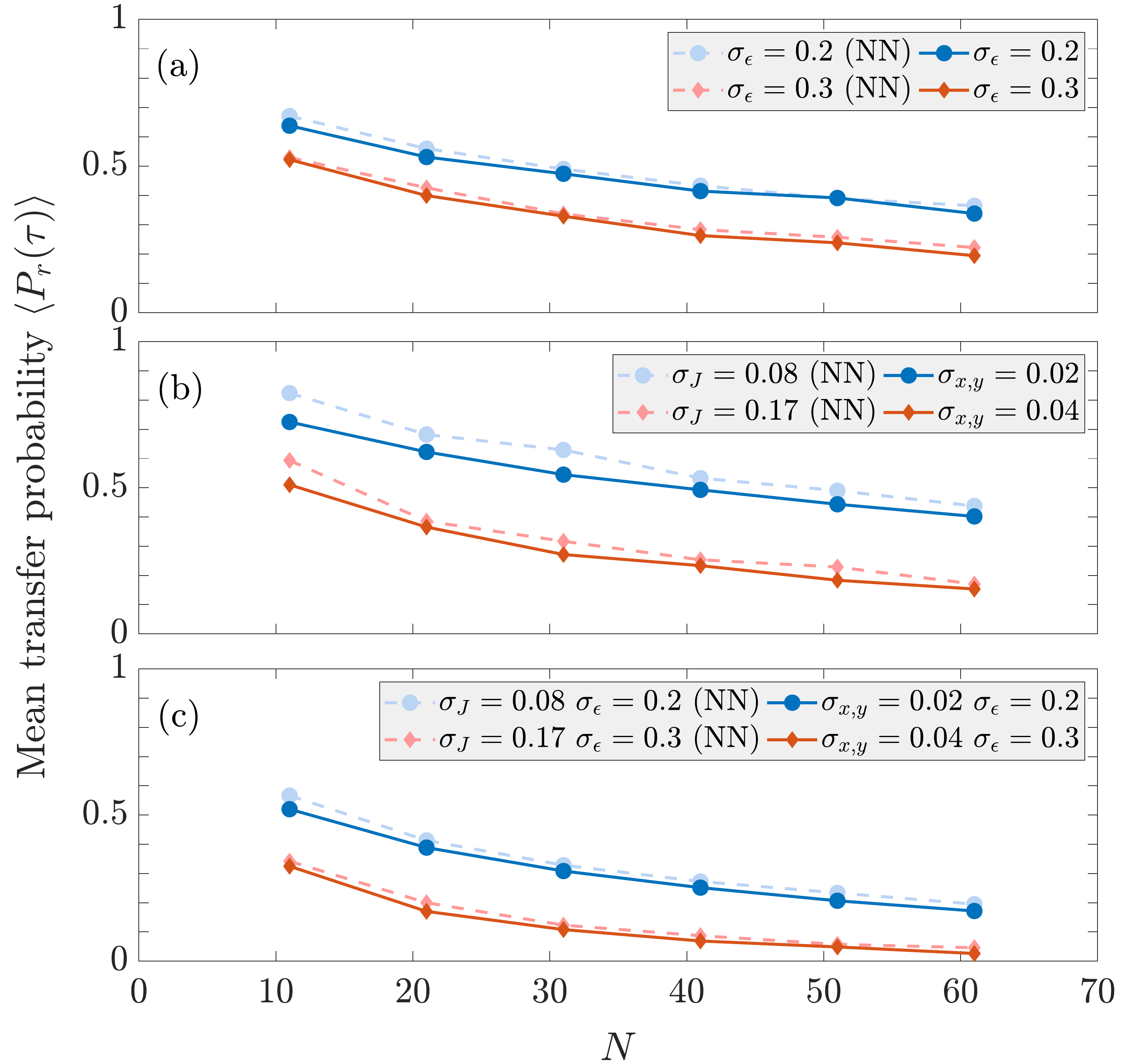}
\caption{Mean excitation transfer probability $\langle P_{r} \rangle$ vs chain length $N$ obtained upon 
averaging over $1000$ independent realizations of disordered chains with long-range interactions (solid lines with filled circles and diamonds) 
and nearest-neighbor interactions (dashed lines with light filled symbols) for 
(a) energy (diagonal) disorder with standard deviations $\sigma_{\epsilon}$,
(b) position (off-diagonal) disorder with standard deviation $\sigma_{x,y}$ or $\sigma_{J}$, and
(c) combination of energy and position disorder.
We use the static couplings of the sender and receiver spins $J_{s,r}=0.49 J/ \sqrt{N}$ having energies $\epsilon_{s,r} =-0.22 J$
($\epsilon_{s,r} =0$ for the nearest-neighbor interacting chains), 
and the evolution is terminated at $t=\tau$ of Fig.~\ref{fig:sttrO}(b).}
\label{fig:SSTDO}
\end{figure}

\paragraph{Static coupling to the chain.}
We first consider the static transfer protocol of  Fig.~\ref{fig:sttrO} with fixed coupling rates $J_{s,r} \simeq 0.5 J/ \sqrt{N}$  
of the sender and receiver spins having energies $\epsilon_{s,r} =-0.22 J$. With the excitation initially localized at the sender spin,
we terminate the evolution when the excitation probability of the receiver spin attains its first maximum at $t=\tau$ of Fig.~\ref{fig:sttrO}(b). 
In Fig.~\ref{fig:SSTDO} we show the transfer probabilities $\langle P_{r} \rangle$ averaged over many independent realizations of disordered
spin chains, involving spin-energy (diagonal) disorder, spin-position (off-diagonal) disorder, and the combination of the two.
As expected, increasing the chain length $N$ decreases the transfer probability which is due to the stronger disorder-induced localization 
of the eigenstates of the chain in the middle of the energy spectrum. We also observe that chains with only the nearest-neighbor interaction
(with $\epsilon_{s,r} =0$) lead to better transfer probability, especially for the case of off-diagonal disorder, Fig.~\ref{fig:SSTDO}(b), 
which is consistent with their larger localization length under otherwise similar conditions, as discussed in Sec.~\ref{sec:loc} and seen in Fig.~\ref{fig:sn}(b).

\begin{figure}[t]
\includegraphics[width=0.9\columnwidth]{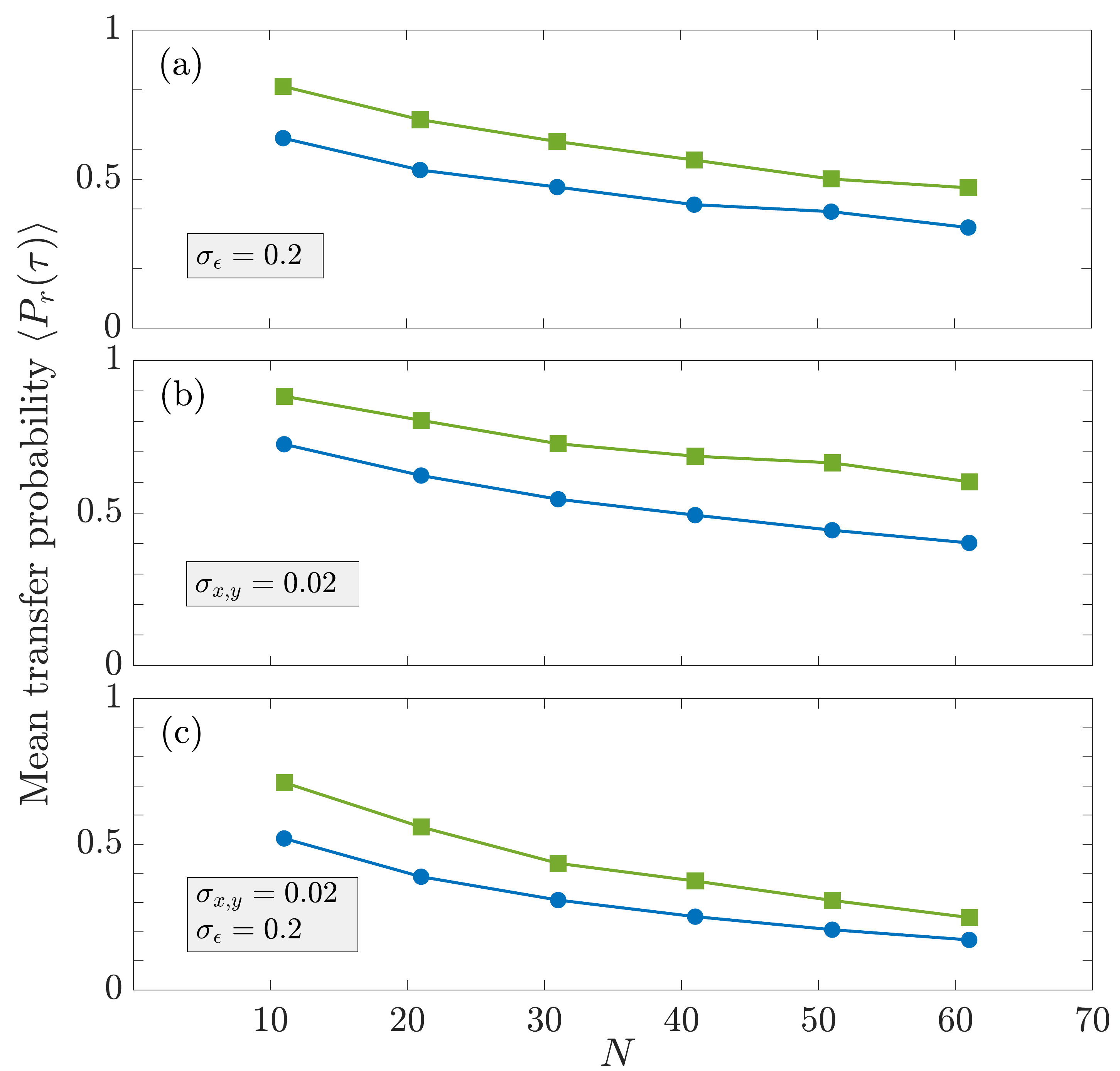}
\caption{Mean stimulated adiabatic excitation transfer probability $\langle P_{r} \rangle$ vs chain length $N$ obtained upon averaging over $1000$ independent realizations of disordered chains with long-range interactions (green solid lines with filled squares), compared to the static transfer of Fig.~\ref{fig:SSTDO} (blue solid lines with filled circles) for 
(a) energy (diagonal) disorder with standard deviations $\sigma_{\epsilon}$,
(b) position (off-diagonal) disorder with standard deviation $\sigma_{x,y}$, and
(c) combination of energy and position disorder.
We use the time-dependent couplings of Eq.~(\ref{eq:stirapulses}) for the sender and receiver spins having energies $\epsilon_{s,r} =-0.22 J$, with the transfer duration $\tau$ of Fig.~\ref{fig:stirapO}(b).}
\label{fig:STIRAPDO}
\end{figure}

\paragraph{Time-dependent adiabatic couplings.}
We finally consider the adiabatic transfer protocol of Fig.~\ref{fig:stirapO} with the time-dependent coupling rates of Eq.~(\ref{eq:stirapulses}) applied to the sender and receiver spins in a counterintuitive order. In Fig. \ref{fig:STIRAPDO} we show the results of our numerical simulations for the transfer probabilities $\langle P_{r} \rangle$ averaged over many independent realizations of disordered spin chains.
Compared to the static transfer protocol, the performance of adiabatic transfer is significantly better for all chain lengths and any kind of disorder, be it diagonal, off-diagonal, or combination of both. We emphasize that in this study, we have focused on the spin excitation or polarization transfer probability. 
In contrast, coherent quantum state transfer is much more sensitive to diagonal disorder leading to larger dephasing during adiabatic transfer that is necessarily slower than the static transfer \cite{petrosyan2010state}.

\section{Conclusions} 
\label{sec:conc}

We have presented the results of our studies of disordered, one-dimensional, long-range interacting spin chains 
and their ability to transfer spin excitation or polarization over long distances. 
We have performed detailed numerical investigations of the localization length in spin chains with either or both diagonal and off-diagonal disorder.
Many of our results concur with the previously known and well-understood properties of disordered spin chains, but we have also encountered interesting manifestations of (de)localization of energy eigenstates that, to the best of our knowledge, have not been properly addressed before in the context of resonant dipole-dipole ($1/r^3$) interactions, and thus may warrant further investigation. 
These, in particular, include delocalization of the eigenstates at the upper edge of the shifted energy band in long-range interacting spin chains with off-diagonal disorder, and the modification of the shifted zero-energy Dyson peak of localization length, which we found to be the most suitable eigenstate for the excitation transfer between the two ends of the chain. 

We have put forward two excitation transfer protocols: a) static protocol involving selective coupling of the sender and receiver spins to the suitable eigenstate of the chain, and b) time-dependent adiabatic protocol involving counter-intuitive sequence of couplings of the sender and receiver spins to the chain, inspired by stimulated Raman adiabatic passage technique widely used in atomic and molecular physics. We have found that the adiabatic transfer of excitation via disordered spin chains has much better performance for all chain length and any kind of disorder, be it diagonal, off-diagonal, or combination of both. This attests, once again, the usefulness of this universal method.

\acknowledgments 
We thank Ivan Khaymovich for helpful comments and suggestions. 
This work was supported by the EU QuantERA Project PACE-IN. 
G. K. acknowledges also the support of PATHOS (FET Open) and DFG (FOR 7274).

\end{document}